\documentclass[preprint,prc,showpacs,preprintnumbers,twocloumn,superscriptaddress,floatfix]{revtex4}

\usepackage{bm}
\usepackage{graphicx,epsfig,latexsym,amssymb}
\usepackage[compact,tiny]{titlesec}
\usepackage{multirow,amsmath}
\usepackage{dcolumn,fancyhdr,makeidx}
\usepackage{subfigure,color}

\begin{document}

\title{Constrained relativistic mean field approach with fixed configurations}

\author{H. F. L\"{u} }
\affiliation{School of Physics, Peking University, Beijing 100871}
\affiliation{School of Science, Chinese Agriculture University,
Beijing 100083}
\author{L. S. Geng}
\affiliation{School of Physics, Peking University, Beijing 100871}
\author{J. Meng}\thanks{e-mail: mengj@pku.edu.cn}
\affiliation{School of Physics, Peking University, Beijing 100871}
\affiliation{Institute of Theoretical Physics, Chinese Academy of
Science, Beijing 100080}
\affiliation{Center of Theoretical Nuclear Physics, National Laboratory of \\
       Heavy Ion Accelerator, Lanzhou 730000}

\begin{abstract}

A diabatic (configuration-fixed) constrained approach to calculate
the potential energy surface (PES) of the nucleus is developed in
the relativistic mean field model.  {As an example}, the potential
energy surfaces of $^{208}$Pb obtained from both adiabatic and
diabatic constrained approaches are investigated and compared. {It
is shown that} the diabatic constrained approach enables one to
decompose the segmented PES obtained in usual adiabatic approaches
into separate parts uniquely characterized by different
configurations, {to follow the evolution of single-particle orbits
till very deformed region}, and to obtain several well defined
deformed excited states which can hardly be expected from the
adiabatic PES's.
\end{abstract}

\pacs{21.10.Dr, 21.10.Re, 21.60.Jz, 21.10.Pc, 21.10.Gv}

 \maketitle

\section{Introduction}

The relativistic mean field (RMF) theory is one of the most
successful microscopic models in nuclear physics. The RMF theory
incorporates from the beginning very important relativistic
effects, such as the existence of two types of potentials (Lorentz
scalar and four-vector) and the resulting strong spin-orbit
interaction, a new saturation mechanism by the relativistic
quenching of the attractive scalar field, and the existence of
anti-particle solutions. The Lorentz covariance and special
relativity make the RMF theory more appealing for studies of high
density and high temperature nuclear
matter~\cite{Serot86,Reinhard89,Ring96}. It has achieved  success
in describing many nuclear phenomena for stable
nuclei~\cite{Reinhard89,Ring96}, exotic
nuclei~\cite{meng96prl,meng98prl} as well as supernova and neutron
stars~\cite{Glendenning}. The RMF theory provides a new
explanation for the identical bands in superdeformed
nuclei~\cite{konig93} and the neutron halo~\cite{meng96prl};
predicts a new phenomenon --- giant neutron halos in heavy nuclei
close to the neutron drip line~\cite{meng98prl,Meng02}; gives
naturally the spin-orbit potential, the origin of the pseudospin
symmetry~\cite{Arima69,Hecht69} as a relativistic
symmetry~\cite{Ginocchio97,meng98prc,meng99prc}, and spin symmetry
in the anti-nucleon spectrum~\cite{Zhou03prl}, and well describes
the magnetic rotation~\cite{Mad00}, the collective multipole
excitations~\cite{Ma02}, and the properties of hypernuclei, etc.
Lately, the ground-state properties of over 7000 nuclei have been
calculated in the RMF+BCS model and good agreements with existing
experimental data are obtained~\cite{Geng05PTP}. Recent and more
complete reviews of the applications of the RMF model,
particularly those on exotic nuclei, can be found in Refs.
\cite{Ring05,meng05ppnp}.

In order to describe the shape of the atomic nucleus and
understand the fusion and fission processes, it is crucial to
obtain the potential energy surface (PES) of the nucleus as a
function of the deformation \cite{Flocard74}. In phenomenological
methods, the PES is obtained by minimizing the total energy of the
system with some shape parameters. In microscopic models, such as
Skryme Hartree Fock and RMF model, PES can be obtained in a
constrained calculation. There are two different ways to obtain
the PES, i.e., adiabatic and configuration-fixed (diabatic)
constrained approaches. In the adiabatic approach, one always
occupies the orbital that results in the lowest configuration
energy. However, the PES obtained in adiabatic calculations may
correspond to different configurations wherever there is orbital
crossing. To obtain the PES for a given configuration, the concept
of the so-called ``parallel transport" should be used, which
enables one to decompose the ground state PES into separate parts
uniquely characterized by the quantum numbers of the occupied
orbits~\cite{Bengtsson89,Meng94}. {This configuration-fixed
constrained approach is often referred to as the diabatic
constrained approach in the literature
~\cite{Bengtsson89,Meng94,Xu98}.}

The diabatic constrained approach is useful to investigate level
crossings which lead to flip-flop situations. It is also useful in
the Generator-Coordinate-Method (GCM) calculations in cases that
excited configurations are included.  Although the so-obtained
excited states cannot always be identified as physical states
since the solutions of the corresponding equations violate the
variational principle, the diabatic constrained approach may serve
as a basis to qualitatively understand the excited states and the
interplay between the excited states and the ground state. To
describe these more quantitatively, one certainly should go beyond
the mean-field level.

Conventionally, the diabatic constrained method has been combined
with the non-relativistic approaches, such as shell model, Skyrme
or Gogny Hartree-Fock or Hartree-Fock-Bogoliubov methods, in
understanding the properties of atomic nuclei, including high-spin
states~\cite{Xu98}, level crossings~\cite{Guo04}, fusion and
fission processes~\cite{DT05}, etc. For example, the interesting
phenomenon of shape coexistence \cite{Cwiok05} can be studied
using PES in complementary to the conventional interpretation of
many particle many hole excitations \cite{May77}. In the RMF
model, the diabatic constrained method has not received enough
attention compared with that of the adiabatic constrained method
and the non-relativistic ones. Therefore, the purpose of this
paper is to develop the diabatic constrained approach within the
RMF model and the doubly magic nucleus $^{208}$Pb is chosen as an
example for demonstration.

This paper is organized as follows: Section II contains an outline
of the RMF model. The adiabatic and diabatic constrained PES in
$^{208}$Pb, together with the diabatic single particle spectra of
the spherical ground state, are analyzed in Section III. The
results are briefly summarized in Section IV.

\section{Theoretical framework}

The relativistic mean field model describes the nucleus as a
system of Dirac nucleons which interact in a relativistic
covariant manner via the meson fields. The meson fields considered
in most applications of the RMF model are the scalar-isoscalar
sigma meson, vector-isoscalar omega meson, vector-isovector rho
meson and the photon: the sigma meson provides the long-range
attractive interaction, the omega meson provides the short and mid
range repulsive interaction, the rho meson is responsible for the
isospin dependence of the nuclear force while the photon accounts
for the electromagnetic interaction. The Lagrangian density used
in our present calculation is of the following form:
 \begin{eqnarray}
 \nonumber
 {\cal L} & = & \bar\psi\left(\rlap{/}p - g_\omega\rlap{/}\omega
               - g_\rho\rlap{/}\vec\rho\vec\tau - \frac{1}{2}e(1 -\tau_3)\rlap{\,/}A
               - g_\sigma\sigma - m\right)\psi
          + \frac{1}{2}\partial_\mu\sigma \partial^\mu \sigma -\frac{1}{2}m_\sigma^2 \sigma^2
               - \frac{1}{3}g_2 \sigma^3 -\frac{1}{4}g_3 \sigma^4
        \\
         & &
           - \frac{1}{4}{\Omega}_{\mu\nu}{\Omega}^{\mu\nu}
               + \frac{1}{2} m_{\omega}^2{ {\bf\omega}_\mu {\bf\omega}^\mu}
               + \frac{1}{4}c_3 {(\omega_\mu \omega^\mu)}^2
            - \frac{1}{4}\vec R_{\mu\nu}\vec R^{\mu\nu}
               + \frac{1}{2} m^2_\rho\vec\rho_\mu\vec\rho^\mu - \frac{1}{4} F_{\mu\nu}
               F^{\mu\nu},
 \label{Eq:lagrangian}
 \end{eqnarray}
where $\psi$ is the Dirac spinor and $\bar\psi=\psi^\dag\gamma^0$.
$m$, $m_{\sigma}$, $m_{\omega}$, and $m_{\rho}$ are the nucleon,
$\sigma$, $\omega$ and $\rho$ meson masses respectively, while
$g_{\sigma}$, $g_{2}$, $g_{3}$, $g_{\omega}$, $c_{3}$, $g_{\rho}$,
 and $e^2/4\pi$ = 1/137 are the corresponding coupling
constants for the mesons and the photon. The field tensors of the
vector mesons ($\omega$ and $\rho$) and of the electromagnetic
fields take the following form:
 \begin{subequations}
  \begin{eqnarray}
   \Omega^{\mu\nu}
     &=& \partial^\mu\omega^\nu - \partial^\nu\omega^\mu   ,\\
   \vec R^{\mu\nu}
     &=& \partial^\mu \vec\rho^\nu - \partial^\nu \vec\rho^\mu
       - 2g_\rho\vec\rho^\mu\times\vec\rho^\nu ,\\
   F^{\mu\nu}
     &=& \partial^{\mu}A^{\nu} - \partial^{\nu}A^{\mu}.
  \end{eqnarray}
 \end{subequations}
With the classical variational principle, one can obtain the
coupled equations of motion, i.e., the Dirac equation for the
nucleons and the Klein-Gordon type equations for the mesons and
the photon. In the most general case, these equations are very
difficult to solve; therefore various symmetries must be utilized
to simplify them.

In the case of an axially symmetric system, the projection of
angular momentum on the z-axis, $\Omega$, and the parity, $\pi$, are
good quantum numbers. Here, we consider only even-even nuclei,
therefore time reversal symmetry is conserved. To solve the RMF
equations for an axially symmetric system, we expand the Dirac
spinors and the Boson fields with the eigenfunctions of a deformed
harmonic oscillator potential in cylindrical coordinates
~\cite{Ring97}, e.g.,
\begin{equation}
 \psi_i (\vec r, t) =
  \left( \begin{array}{c}   f_i (\vec r) \\ ig_i (\vec r)
  \end{array} \right) \chi_{t_i} (t) =
   \frac{\chi_{t_i} (t)}{\sqrt{2\pi}} \left( \begin{array}{c}
    f_i^+ (z,r_\bot) e^{i(\Omega_i -1/2)\theta} \\
      f_i^- (z,r_\bot) e^{i(\Omega_i +1/2)\theta} \\
        ig_i^+ (z,r_\bot) e^{i(\Omega_i -1/2)\theta} \\
         ig_i^- (z,r_\bot) e^{i(\Omega_i +1/2)\theta}
          \end{array} \right).
\label{spinor}
\end{equation}
 20 oscillator shells have been used to expand
both the Fermion fields and Boson fields to guarantee convergence.

The potential energy surface is obtained through the constrained
calculation. More specifically, the binding energy at certain
deformation is obtained by constraining the quadruple moment
$\langle \hat Q_2 \rangle$ to a given value $\mu$, i.e.,
\begin{equation}
\langle H' \rangle   =  \langle H \rangle +\frac{1}{2} C( \langle
\hat{Q}_2 \rangle - \mu )^2 ,
\end{equation}
with
\begin{equation}
\langle \hat Q_2 \rangle = \langle \hat Q_2 \rangle_n +\langle
\hat Q_2 \rangle_p ,
\end{equation}
where $\langle \hat Q_2 \rangle_{n,p}  = \langle 2r^2 P_2(cos
\theta) \rangle_{n,p}
    = \langle 2z^2 -x^2 -y^2 \rangle_{n,p}$ and $C$ is the curvature constant parameter.
The more often used deformation parameter $\beta_2$ is related with
the expectation value $\langle \hat Q_2 \rangle $ by $\displaystyle
\langle \hat Q_2 \rangle = \frac{3}{\sqrt{5\pi}} A r^2 \beta_2$, $r=
R_0 A^{1/3}~~ (R_0 = 1.2$~fm) and $A$ is the mass number. By varying
$\mu$, the binding energy at different deformation can be obtained.
 {In principle, one has to follow a multidimensional energy surface.
However, for the sake of efficiency,  a one-parameter line, e.g.
$\beta_2$, is often used in microscopic constrained RMF
calculations due to the self-consistency~\cite{Ring80} and the
small effects of other deformations, such as the hexadecupole
deformation.}

In order to obtain the diabatic PES with a fixed configuration,
the occupied orbits are determined by the so-called ``parallel
transport", i.e.,
\begin{equation}
  \langle \psi_i(q) |\psi_j (q+\Delta q) \rangle|_{\Delta q \to
  0} \approx \delta_{ij},
  \label{condition}
   \end{equation}
where $i$ and $j$  enumerate all the single-particle states of two
adjacent configurations. In such a way, the original configuration
at $q$ can be tracked and the corresponding PES can be obtained as
a function of the deformation \cite{Guo05}.

{In principle, if $\Delta q$ is small
enough, the configuration at $q+\Delta q$ should be the same as
that at $q$. However, due to numerical difficulties, condition in
Eq.~(\ref{condition}) can not be rigorously implemented.
Therefore, the following procedures are adopted in our study:
First, the wavefunction and occupation number of every state at
the initial configuration $q$ are recorded. Second, for each state
i at configuration $q$, we search all the states of configuration
$q+\Delta q$ for state j that has the largest overlap with state
i. It is to be noted that state i and state j have the same
quantum numbers $\Omega^\pi$. Once state j is determined, the
occupation number of state i will be transferred to state j. This
procedure is repeated till the occupation number of each state at
configuration $q+\Delta q$ fixed. In case that many single
particle states with the same symmetry are very close to each
other, we adjust the step size of $\Delta q$ to avoid the mismatch
of the single-particle configuration. }

\section{Results and Discussion}

In the preceding section, we described how to implement the diabatic
constrained method within the relativistic mean field model. In this
section, we would like to compare the results obtained with this
method and those with the usual adiabatic constrained one. For this
purpose, we study the potential energy surface of $^{208}$Pb. This
nucleus is chosen due to the following considerations: First, its
properties can be well described by the RMF model; second, as a
doubly magic nucleus, the pairing correlation can be safely ignored.
The latter is particularly important because the pairing correlation
can complicate the issue dramatically~\cite{Guo04}.
{Taking into account the pairing correlation, such as using a
constant gap pairing scheme, does not qualitatively alter our
results and the corresponding conclusions. However, in the presence
of the pairing correlation, due to the fractioned occupation number,
it is not convenient to track configuration changes and discuss the
relevant physics. Therefore, only the results without the pairing
correlation are presented.} It should be mentioned that the
parameter set used is PK1, which has been carefully adjusted to
reproduce both the nuclear matter saturation properties and the
ground-state properties of finite nuclei with a microscopic recipe
for the center of mass correction~\cite{Long04}.

\begin{figure}
 \centerline{\epsfig{figure=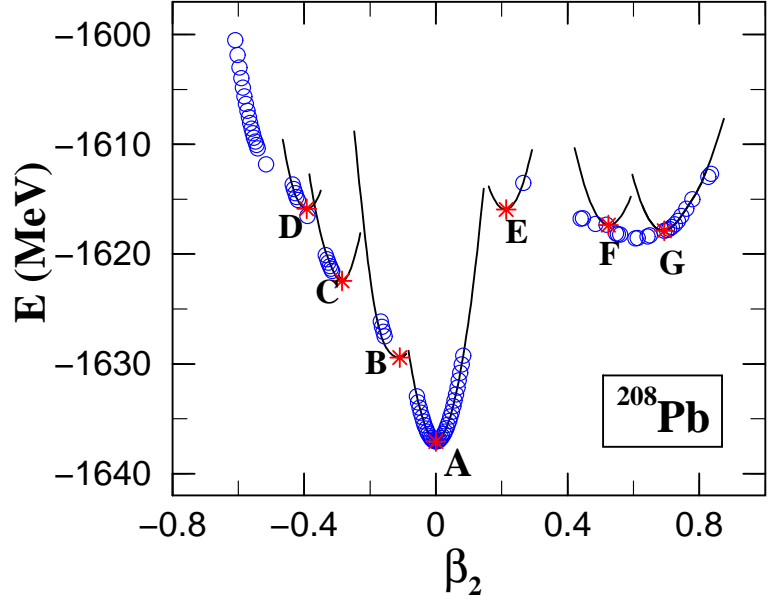,width=10.cm}}
  \caption{(Color online) Potential energy surfaces of $^{208}$Pb obtained in adiabatic
     (open circles) and diabatic calculations (solid lines).
    The local minima on the corresponding diabatic PES are denoted
     as stars and labelled by different capital letters, whose configuration, binding energy and deformation
    are tabulated in Table.~\ref{contab}.}
 \label{PES}
\end{figure}

In Fig.~\ref{PES}, the PES of $^{208}$Pb obtained from adiabatic
(open circle) and diabatic (solid line) calculations are plotted
as functions of the deformation parameter $\beta_2$. Surprisingly,
it is found that even for $^{208}$Pb, which has a rigid spherical
shape at ground state, the adiabatic PES is somewhat complicated,
which can be easily seen from the many broken regions on the PES.
The fact that there exists un-converged region on the adiabatic
PES has been known for a long time. Over the years, it has been
argued that it might originate from (i) the abrupt change of
configuration at a certain point due to the no crossing
rule~\cite{Ring80}; (ii) change of the mean fields due to varied
configuration~\cite{Guo05}; and/or (iii) the increased mixing of
all low-lying states due to their near degeneracy~\cite{Bender04}.
On the other hand, the diabatic method is known to be able to
connect broken regions existing in the adiabatic method. This can
be clearly seen in Fig.~\ref{PES}. In addition, from
Fig.~\ref{PES}, as expected, one can see that the diabatic results
(solid lines) can reproduce the adiabatic results (open circles)
very well if they both exist at the same region.

It should be noted that several local minima which do not exist on
the adiabatic PES appear on the diabatic one. These minima, which
could be viewed as excited states, are denoted as stars and
labelled by capital letters in Fig.~\ref{PES}.  Since in the
diabatic method, the single-particle levels of each configuration
can be unambiguously determined (see the preceding section), we
tabulate the corresponding configuration of each new local minimum
in Table~\ref{contab}. Each minimum is labelled by a bracket
($E_x,\beta_2$), where $E_x$ is the excited energy relative to the
ground state {\bf A} and $\beta_2$ is the corresponding
deformation parameter. The configuration of each state is given
with respect to its core listed in the second row of the table.
The configuration of state {\bf F} relative to that of state {\bf
E} is not given simply because the difference between these two
configurations is too large to fit into the table. We would like
to point out a few more interesting things:
\begin{itemize}

\item From Table~\ref{contab}, it is seen that for most cases both
neutrons and protons are excited as configuration changes from one
to the neighboring excited state. This might be related to the
strong interaction between neutrons and protons. In this sense, a
self-consistent calculation, such as the present one, is important
to take care of this interaction and yields reliable results.

\item Due to the fact that different local minima have different
configuration, even the barrier between them is very small, they
can still be well defined, such as $\mathbf{B}$ and $\mathbf{A}$,
$\mathbf{F}$ and $\mathbf{G}$~\cite{Bengtsson89}.

\end{itemize}

In Fig.1,  the diabatic potential (solid curve) in Fig. 1 show
discontinuity around the deformations $\beta \sim 0.4$, which is due
to the absence of the other diabatic curves with different
configurations. If more careful diabatic constrained calculations
with these configurations were done, the diabatic curves will become
continuous and the two values at some quadrupole deformations, e.g.,
$\beta \sim 0.4$, would belong to one or two of these diabatic
curves.

\begin{figure}
 \centering
  \begin{minipage}[t]{.485\textwidth}
  \centering
  \includegraphics[width=1.25\textwidth]{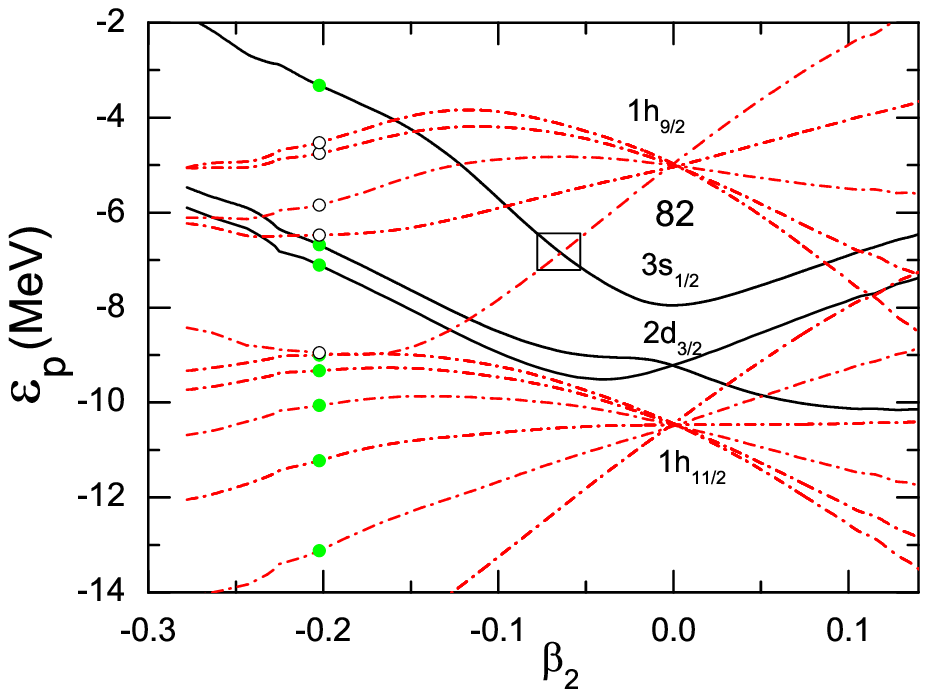}
  \end{minipage}
 \hspace{.5em}
  \begin{minipage}[t]{.485\textwidth}
  \centering
  \includegraphics[width=1.25\textwidth]{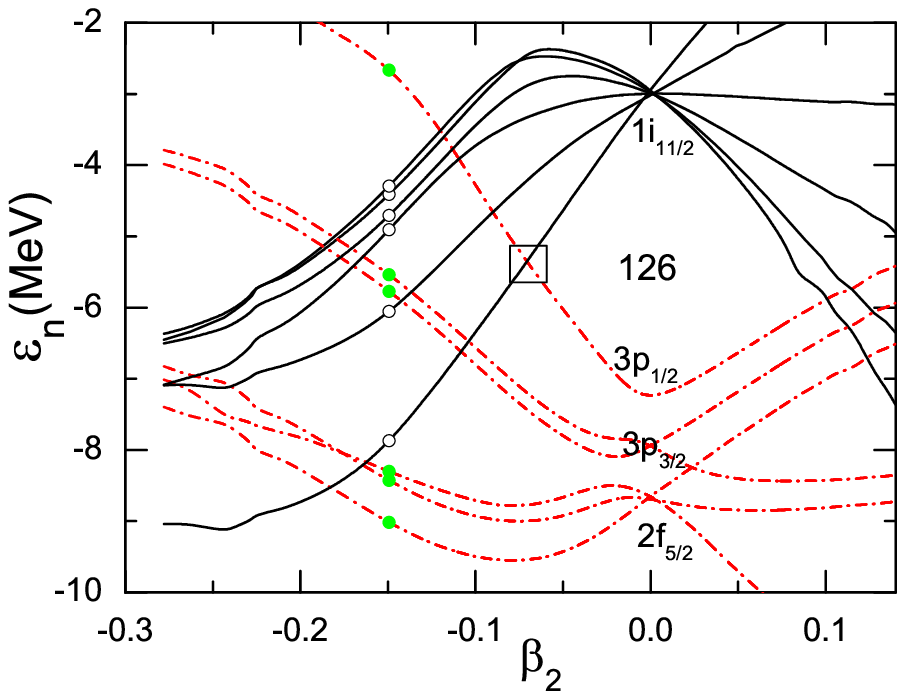}
  \end{minipage}
 \caption{(Color online) Proton (left panel) and neutron (right
panel) single particle states of  $^{208}$Pb obtained in diabatic
calculations as functions of the deformation parameter $\beta_2$.
 Solid line and dot-dashed line represent positive and
negative parity states, respectively. Filled (open) circle denotes
whether the corresponding state is occupied (unoccupied). The
original configuration is that of the ground state. For the sake
of clarity, only those states near the fermi surface are shown;
and the level crossing points are highlighted by rectangles. }
  \label{ad-sp}
\end{figure}

To provide a microscopic explanation of Fig.~\ref{PES}, we plot the
single particle spectra of $^{208}$Pb near Fermi level obtained from
the diabatic calculation in Fig.~\ref{ad-sp}. Solid (dot-dashed)
lines represent positive (negative) parity, filled (open) circles
label the occupied (unoccupied) states, and the original
configuration is that of the spherical case {\bf A} (see
Fig.~\ref{PES}). The level crossing regions are highlighted by
rectangles. From the left panel of Fig.~\ref{ad-sp}, one can easily
see those high $j$ levels originating from $1h_{9/2}$ come down
while those low $j$ levels, such as $3s_{1/2}$ and $2d_{3/2}$, go up
as the nucleus becomes more oblate. At around $\beta_2\approx -0.1$,
the first level crossing happens between $3s_{1/2}$ and one
$1h_{9/2}$ state. The same is true for neutrons, in particular, the
first level crossing happens between $3p_{1/2}$ and one $1i_{11/2}$
state at $|\beta_2|$ slightly larger than that where the first
proton level crossing occurs. It shows that a diabatic calculation
is important to obtain the smooth evolution of the single-particle
levels as a function of deformation. Otherwise irregularity of the
single-particle levels will appear due to the occupation of the
lower orbits in the adiabatic calculation.  We should point out that
here proton and neutron level crossings occur almost simultaneously
mainly due to the fact that their corresponding Fermi levels are
close to each other in the case of $^{208}$Pb. It may not be the
case for either proton or neutron rich nuclei where the proton and
neutron Fermi surfaces are quite different. Beyond the crossing
points, high $j$ levels continue coming down while low $j$ levels
continue going up, which results in a very steep potential energy
surface. This is the case for most diabatic PES's displayed in
Fig.~\ref{PES}.

\begin{figure}
 \centering
 \begin{minipage}[t]{.485\textwidth}
  \centering
  \includegraphics[width=1.25\textwidth]{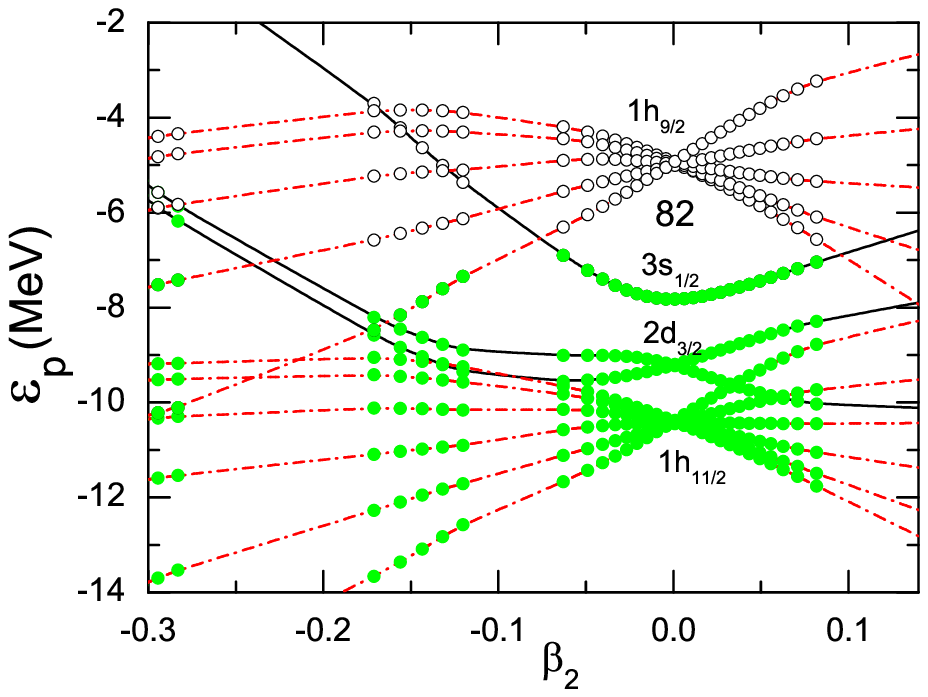}
  \end{minipage}
 \hspace{.5em}
  \begin{minipage}[t]{.485\textwidth}
  \centering
  \includegraphics[width=1.25\textwidth]{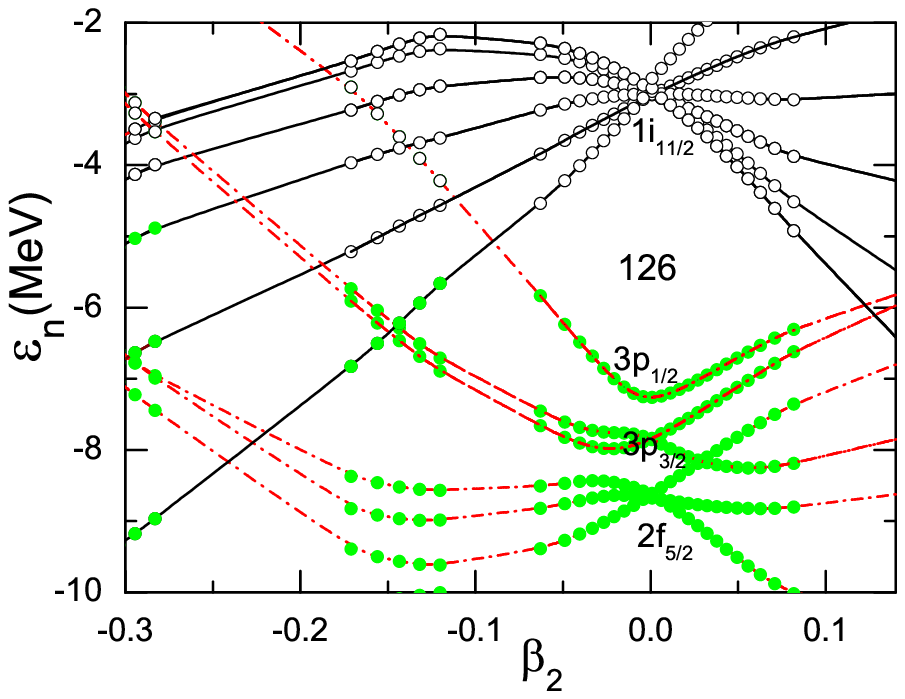}
  \end{minipage}
 \caption{(Color online) Proton (left panel) and neutron (right
panel) single particle states of  $^{208}$Pb obtained in adiabatic
calculations as functions of the deformation parameter $\beta_2$.}
  \label{di-sp}
\end{figure}

{Finally, we plot the adiabatic single
particle levels in Fig.~\ref{di-sp}. They look similar to those
shown in Fig.~\ref{ad-sp} but differ in two aspects. First, the
diabatic single particle levels are continuously distributed as
functions of $\beta_2$ but the adiabatic single particle levels
are not. Their broken pattern is the same as that of the adiabatic
PES shown in Fig.~\ref{PES}. The underlying reason is, as we have
mentioned above, that the sudden jump of the adiabatic PES is
mostly due to the abrupt change of the corresponding configuration
\cite{Guo05}. For example, the crossing of $3p_{1/2}$ and
$1i_{11/2}$ at $\beta_2\approx -0.1$ results in the broken PES
shown in Fig.~\ref{PES}. Second, the diabatic single particle
levels are occupied according to the dictated configuration but
the adiabatic single particle levels are so occupied to give the
largest binding energy.}

\section{Summary}

A diabatic constrained relativistic mean field approach is
proposed to calculate the potential energy surface of the nucleus
and applied to $^{208}$Pb. Although both adiabatic and diabatic
constrained calculations yield almost the same ground state PES,
the diabatic one has the advantages that it enables one (i) to
decompose the segmented PES obtained in usual adiabatic approaches
into separate parts characterized uniquely by different
configurations; and (ii) to define the single particle orbits at
each deformation by their quantum numbers unambiguously determined
from their counterparts at the spherical configuration. Thus,
physics behind the adiabatic PES can be understood more clearly in
the diabatic picture.

\vspace{2em}This work is partly supported by the National Natural
Science Foundation of China under Grant No. 10435010, 10221003,
and 10447101, the Doctoral Program Foundation from the Ministry of
Education in China.


\begin{table}
  \caption{Configurations of the five local minima ({\bf B, C, D, E, G}) on the diabatic
   potential energy surfaces of $^{208}$Pb (see also Fig.~\ref{PES}). Each minimum is labelled
   by a bracket ($E_x,\beta_2$) where $E_x$ is the excited energy relative to the ground state {\bf A} and
    $\beta_2$, the corresponding deformation parameter. The configuration of each
   state is relative to the core as listed in the second column.}
  \begin{tabular}{c|c|c|c}
  \hline\hline
                         &               & \multicolumn{2}{c}{Configuration}      \\ \cline{3-4}
      \raisebox{2ex}[0pt]{State}             &\raisebox{2ex}[0pt]{Core}        &   $\pi$  &   $\nu $  \\ \hline
  B~~ (7.63,-0.11)  &A   &  $  (3s_{1/2})^{-1}(1h_{9/2})^1 $ &   $      (3p_{1/2})^{-1}(1i_{11/2})^1  $
                       \\  \hline
  C~~ (15.63,-0.28) &B &  $   (2d_{3/2})^{-2} (1i_{13/2})^1  (2f_{7/2})^1    $&
                       $     (3p_{3/2})^{-2}  (2g_{9/2})^1 (1j_{15/2})^1 $
                          \\  \hline
  D~~ (21.19,-0.39) &C   &    &
               $   (2f_{5/2})^{-3} (2h_{9/2})^2 (1j_{13/2})^1     $
                          \\   \hline\hline
  E~~ (21.13,0.21)  &A  &  $   (3s_{1/2})^{-1} (2d_{3/2})^{-1} (1h_{9/2})^2  $  &
              $      (3p_{1/2})^{-1} (2f_{5/2})^{-2} (2g_{9/2})^2 (1i_{11/2})^1 $
                         \\    \hline
  G~~ (19.19,0.69)  &F  &  $  (1g_{7/2})^{-1} (1h_{11/2})^{-1} (1h_{9/2})^1 (1i_{13/2})^1   $ &
                   $     (3p_{3/2})^{-1} (2f_{7/2})^{-1} (1i_{11/2})^1 (1j_{15/2})^1  $
                           \\ \hline\hline
  \end{tabular}
  \label{contab}
 \end{table}


\end{document}